\documentclass[runningheads]{llncs}

\usepackage{amsfonts}
\usepackage{fixmath}
\usepackage{setspace}
\usepackage{enumitem}
\usepackage{amsmath}
\usepackage{algorithmic}
\usepackage{algorithm,xcolor,booktabs}

\newcounter{myprot}
\newcounter{myalg}
\newcounter{myexam}
\newcounter{mythm}
\newcounter{mylem}
\newcounter{mycor}
\newcounter{myobs}
\newcounter{mydef}
\newenvironment{mydef}
{\refstepcounter{mydef} \vspace{1em} \noindent \textbf{DEFINITION \arabic{mydef}:}}
{\vspace{.5em}}

\newcounter{myconj}
\usepackage{mathtools,url}
\usepackage{latexsym}
\usepackage{subcaption}  
\usepackage[font={small,sf}]{caption}

\makeatletter
\g@addto@macro{\UrlBreaks}{\UrlOrds}
\makeatother

\renewcommand{\footnotesize}{\fontsize{8pt}{10pt}\selectfont}

\newif\ifinappendix
\let\oldappendix\appendix
\renewcommand{\appendix}{
  \oldappendix
  \inappendixtrue
}

\usepackage{booktabs}
\usepackage[hidelinks]{hyperref}
\usepackage{xfrac}
\usepackage{wrapfig,multirow,relsize}
\usepackage[space,compress]{cite}
\usepackage{microtype}

\newcommand{\OmitText}[1]{ {} }

\usepackage{xspace}

\newcommand{\para }[1]{\smallskip \noindent {\bf #1}}
\newcommand{\1}{{\em (i)}}
\newcommand{\2}{{\em (ii)}}
\newcommand{\3}{{\em (iii)}}

\usepackage[all]{hypcap}

\usepackage[capitalise,nameinlink]{cleveref}
\crefname{section}{Sect.}{Sect.}
\Crefname{section}{Section}{Sections}

\usepackage{url}
\makeatletter
\g@addto@macro{\UrlBreaks}{\UrlOrds}
\makeatother
\makeatletter
\def\Url@twoslashes{\mathchar`\/\@ifnextchar/{\kern-.2em}{}}
\g@addto@macro\UrlSpecials{\do\/{\Url@twoslashes}}
\makeatother

\begin{document}

\title{Bonded Mining: Difficulty Adjustment\\
by Miner Commitment}

\author{George Bissias\inst{1} \and
David Thibodeau\inst{2} \and
Brian N. Levine\inst{1}} 

\authorrunning{G. Bissias et al.}

\institute{College of Information and Computer Sciences,  UMass Amherst
\email{\{gbiss,levine\}@cs.umass.edu} \and
Florida Department of Corrections \\
\email{davidpthibodeau@gmail.com }\\
}

\date{}
\maketitle

\setlength{\belowdisplayskip}{3pt} 
\setlength{\belowdisplayshortskip}{3pt}
\setlength{\abovedisplayskip}{3pt} 
\setlength{\abovedisplayshortskip}{3pt}

\begin{abstract}
Proof-of-work blockchains must implement a difficulty adjustment algorithm (DAA) in order to maintain a consistent inter-arrival time between blocks. Conventional DAAs are essentially feedback controllers, and as such, they are inherently reactive. This approach leaves them susceptible to manipulation and often causes them to either under- or over-correct. We present {\em Bonded Mining}, a proactive DAA that works by collecting hash rate commitments secured by bond from miners. The difficulty is set directly from the commitments and the bond is used to penalize miners who deviate from their commitment. We devise a statistical test that is capable of detecting hash rate deviations by utilizing only on-blockchain data. The test is sensitive enough to detect a variety of deviations from commitments, while almost never  misclassifying honest miners. We demonstrate in simulation that, under reasonable assumptions, Bonded Mining is more effective at maintaining a target block time than the Bitcoin Cash DAA, one of the newest and most dynamic DAAs currently deployed. 
In this preliminary work, the lowest hash rate miner our approach supports is 1\% of the total and we directly consider only two types of fundamental attacks. Future work will address these limitations.

\keywords{Difficulty adjustment  \and protocols \and cryptocurrencies}
\end{abstract}

\section{Introduction}
\label{sec:intro}

Blockchain protocols maintain a public ledger of account balances that are updated by authorized transactions. Proof-of-work (PoW) mining is the process of assembling transactions into blocks and earning the right to add the block to a growing chain~\cite{Nakamoto:2009}. PoW mining involves repeatedly cryptographically hashing the assembled block, each time with a different nonce. The hashes are generated uniformly at random from a space with maximum value $S$. When a hash falls below a {\em target} $t$, the corresponding block is said to be mined, and it is added to the blockchain. Closely related to target is the \emph{difficulty}\footnote{Technically $D = t_0 / t$, where $t_0 \approx S$ is the target with highest possible difficulty, but this detail is not important for our analysis.} $D$, which is equal to $S / t$. The expected time required to mine a block is a function of $D$ and the rate that hashes are generated, or \emph{hash rate} $h$. Hash rate  fluctuates (sometimes rapidly) over time, and therefore PoW blockchains must adjust $D$  to ensure that the expected block time remains roughly constant. Currently, all PoW blockchains use a \emph{difficulty adjustment algorithm} (DAA) to adjust $D$ as $h$ fluctuates. 

Although implementations vary widely, each DAA is essentially a feedback controller analogous to a thermostat. The DAA uses previous block creation times to detect a change in $h$, and then it makes an adjustment to $D$ in order to move the expected times toward the desired value $T$. There are three major limitations to this {\em reactive} approach.
\begin{enumerate}[nosep,topsep=3pt]
\item There is a tendency to either over or under correct, which can cause oscillations in block time~\cite{Stone:2017b,Stone:2017}.
\item Contentious hard forks create significant instability in block times for minority hash rate blockchains, which must resort to a backup controller that compensates for swings in miner hash rate allocation preference~\cite{Zegers:2017}. 
\item Most control algorithms can be gamed by miners without consequence  in order to extract higher rewards\cite{Kiraly:2018,Kwon:2019},  causing fluctuations in block time as a result.
\item Feedback control is inherently reactionary; it only uses historical block time and difficulty data to produce future difficulty values.
\end{enumerate}

\para{Contributions.} We present \emph{Bonded Mining}, a protocol that enhances PoW mining with a {\em proactive} approach to difficulty adjustment so that inter-block times are always near their desired value despite sudden hash rate changes.  The idea is to ask miners to \emph{commit} to their individual hash rate and financially bind them to 
it by holding \emph{bond}. Difficulty is adjusted based on these self-reported commitments. 
Miners are incentivized to commit to a realistic estimate of their future hash rate and honor their commitment, even if it becomes nominally more profitable to direct their hash rate elsewhere. The protocol is flexible: miner commitments last  until they mine their next block;  and they can deviate from the commitment (incurring a penalty commensurate with their deviation) as long as they are truthful about the deviation.

For security, we derive a statistical test (using on-blockchain data only)  that is capable of detecting both short- and long-term deception from miners. Miners who fail the test suffer a significant financial penalty. The test is sensitive enough to detect a miner who drops to 20\% of her commitment for a week or more, and it can also detect when she strays from her commitment by as little as 1\% every block over the course of 70 days or more. This sensitivity comes with very little risk for honest miners. Even when a miner deviates from her commitment, if she truthfully reports that deviation, then the probability of failing \emph{any} test over the course of a year is less than 0.3\%.

Because of its proactive design and the penalties associated with deception, Bonded Mining is better capable of maintaining the desired expected block times than are conventional, reactive approaches. In Bonded Mining, the extent to which block times remain close to desired time $T$ is the extent to which miners value their bond more than a change in their hash rate. In simulation, we find that even when miners are willing to sacrifice 25\% of their bond in order to change their hash rate, Bonded Mining still demonstrates lower amplitude and duration of deviations from $T$ than does the DAA of Bitcoin Cash, the latter of which is one of the newest and most dynamic DAAs currently deployed.

\section{Protocol Details}
\label{sec:protocol}
\begin{figure}[t]
   \centering
   \includegraphics[width=0.9\columnwidth]{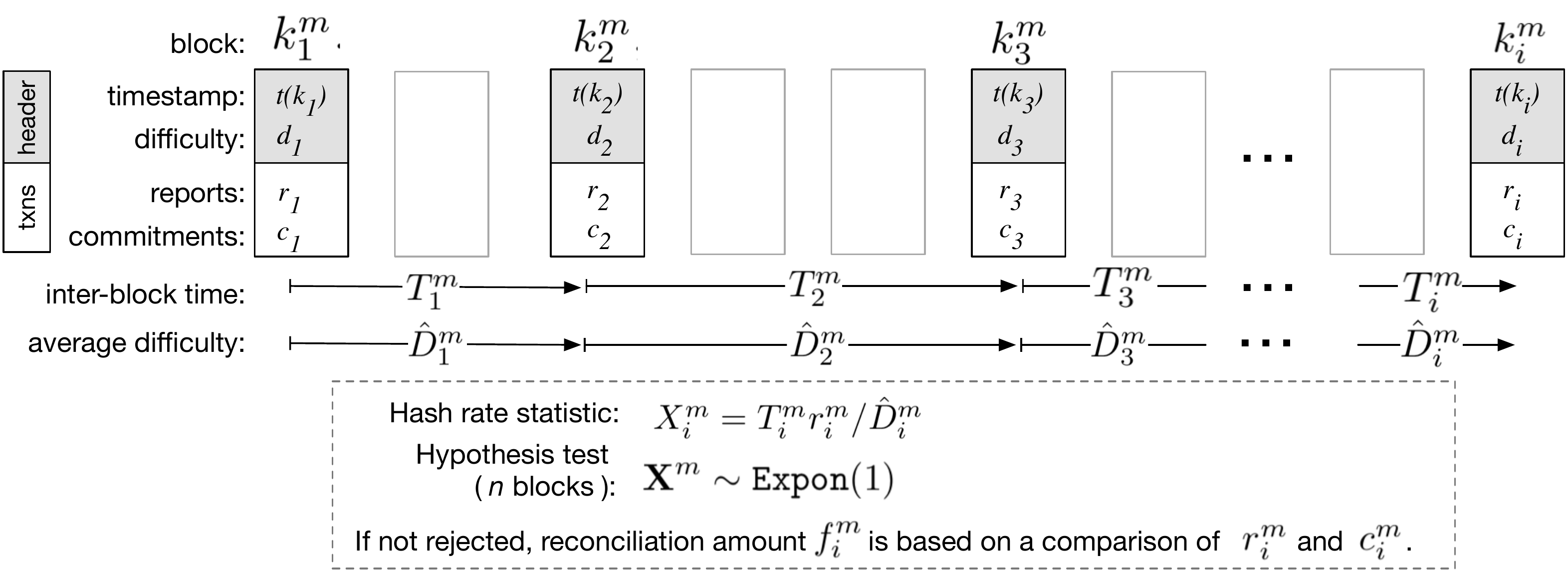} 
   \caption{An overview of Bonded Mining for miner $m$.}
   \label{fig:protocol}
\end{figure}
Bonded Mining enhances existing PoW mining schemes by adding a {\em collateral} requirement to ensure hash rate commitments are honored. Figure~\ref{fig:protocol} illustrates many aspects of the protocol, and a List of Symbols appears in the Appendix. To bootstrap, miner $m$ must post \emph{bond} to the blockchain by paying $b$ coins into a \emph{bond pool} account via a validated \emph{deposit transaction}. The bond is posted prior to mining the first block and is locked until the miner produces his first $n$ blocks, where $n$ is a tunable security parameter depending on the miner's hash rate. As part of the deposit transaction, miner $m$ also stipulates his \emph{commitment} $c_1^m$ for the hash rate he will apply to mining his first block, $k^m_1$.  

\para{Mining.} Now consider the next $n-1$ blocks mined by $m$: $k_2^m, \ldots, k_n^m$. Define $t(k_i^m)$ as the time that $k_i^m$ is mined. Let $t(k_0^m)$ be defined as the time of the block containing the deposit transaction. When mining block $k_i^m$, miner $m$ adds a new deposit transaction to the set of transactions being mined, which deposits an additional $b$ coins to the bond pool. In assembling the block, he uses the typical block header, and includes two new pieces of information: \emph{report} $r_i^m$ and commitment $c_{i+1}^m$. Report $r_i^m$ is an attestation from miner $m$ of his actual average hash rate during the time period between $t(k_{i-1}^m)$ and $t(k_i^m)$. And $c_{i+1}^m$ is  $m$'s commitment for the next block $k_{i+1}^m$. If $m$ wishes to stop after $n$ blocks, then he can issue a divestment transaction (see \emph{Bond state} section below) and $c_{n+1}^m$ is unnecessary; otherwise $c_{n+1}^m$ should contain the commitment for the next block. When block $k_i^m$ is eventually mined by $m$, the coinbase is immediately transferred to the miner's wallet, but bond is released only upon reconciliation. To be clear: the miner can adjust commitments each block starting from block $k_2^m$; but reconciliation begins from block $k_n^m$ onward and evaluates a window of the previous $n$ blocks.

\para{Reconciliation.} If $i$ blocks, $i > n$, have been mined by $m$, then his block $k_i^m$ includes  a \emph{reconciliation transaction} that pays himself a \emph{reconciliation payment} $f_i^m \leq b$ from the bond pool. $f_i^m$
is the refunded portion of the bond deposited by $m$ while mining block $k_{i-n}^m$, which is now eligible to be reconciled. 
(Miners never forfeit coinbase rewards.) Miner $m$ signs and confirms his own reconciliation transaction, but if he repays an inappropriate amount, then the transaction is considered to be invalid by other miners and the entire block is ignored. The value of $f^m_i$ is determined in two stages. The first tests  $m$'s {\em reporting accuracy} via binary hypothesis test $\texttt{Valid}(\mathbf{r}^m)$, which deterministically rejects or accepts the null hypothesis that the inter-block times from $t(k^m_{i-n}), \ldots, t(k^m_i)$ are samples that came from the distribution implied by reports $\mathbf{r}^m = r^m_{i-n}, \ldots, r^m_i$. If the null hypothesis is rejected, then $f_i^m = 0$; i.e., $m$ loses all his bond deposited in block $k_{i-n}^m$. However, if the null hypothesis is not rejected, then in the second stage we evaluate  $m$'s {\em commitment fulfillment} by setting
\begin{equation}
\label{eq:reconcile}
f^m_i = b - b \cdot \pmb{\min} \left\{ 1,  \left| \frac{r^m_i - c^m_i}{c^m_i} \right| \right\}.
\end{equation}
We tune the hypothesis test so that if $m$ honestly reports his hash rate for each block $k_j^m$, then he is very likely to pass, and then $m$ will be repaid his bond as the absolute difference between his committed and actual hash rates. 

\para{Bond State.} We define  four distinct states for bonded miners: \emph{Bootstrapping}, \emph{Fully Bonded}, \emph{Divested}, and \emph{Abandoned}. A new miner who has not yet deposited $n b$ total bond is bootstrapping. Once she has contributed  $n$ mined blocks to the main chain, her total bond reaches $n b$ and the miner is considered fully bonded. Note that the number of blocks required to reach the Fully Bonded state depends on the miner's committed hash rate as a fraction of the total. Accordingly, miners may fluctuate between Bootstrapping and Fully Bonded states if the total hash rate changes. A fully bonded miner is eligible to divest her bond in order to reduce her hash rate on the blockchain to zero. The miner signals this intent by submitting a \emph{divestment transaction}, which can appear in any block, not just one that she mines. The transaction contains reconciliation payments (see above) for each of her $n$ bond deposits that remain in the bond pool. Once the divestment transaction is confirmed, the miner is fully divested: she is committed to zero hash rate and has no remaining bond. In order to begin mining again, a divested miner must proceed through the Bootstrap state. Finally, a miner can reach the Abandoned state from either the Bootstrapping or Fully Bonded states if she fails to generate a block in a reasonable amount of time given her last commitment (see below). If the miner reaches the Abandoned state, all of her bond is lost, and she transitions to the Divested state. (The bond is burned, but could be redistributed as future coinbase.)

\para{Abandonment Detection.} The test for abandonment is distinct from the test $\texttt{Valid}$. It is conducted every block (as opposed to only those generated by the miner). For each miner $m$ in either the Bootstrapping or Fully Bonded state, we have $c^m_i$, the latest commitment from $m$. We argue in Section~\ref{sec:test} that if $m$ honors her commitment, then the inter-arrival time of her $i$th block, $T^m_i$, is exponential. Specifically, $T^m_i \sim \texttt{Expon}(T/c^m_i)$ where $T$ is the target inter-block interval for the network. Now let $Q(p)$ be the quantile function for $T^m_i$ such that $P(T^m_i < Q(p)) < p$. With probability $p$, we know that $T^m_i$ will be less than $Q(p)$. Therefore, for large $p$, we can be almost certain that $m$ has abandoned mining at her committed hash rate if no block hash been seen for time $Q(p)$.

For example, consider a miner who commits to 10\% of the total hash rate in a protocol like Bitcoin with $T=10$ minute block times. With probability exceeding 99.999\%, it will take $m$ no more than 20 hours to mine her next block. Thus, by setting the abandonment time to 20 hours, we can be highly confident that $m$ has in fact abandoned mining at her committed hash rate.

\para{Difficulty Adjustment.} Let $c_i$ and $h_i$ be the total {\em committed} and {\em actual} hash rates, respectively, across all miners for the time period in which block $k_i$ will be mined. Bonded Mining makes the assumption that security parameters $b$ and $n$ are tuned so that miners are incentivized to honor their committed hash rate  during this time period, i.e. $c_i = h_i$. It is known~\cite{Ozisik:july2017} that
the hash target $t$ is related to total hash rate $h_i$ by the equation
\begin{equation}
\label{eq:fundamental}
h_i T = \sfrac{S}{t},
\end{equation}
where $T$ is the expected time to produce a block and $S$ is the size of the hash space, as defined above. Given the equivalences $c_i = h_i$ and $D = S / t$ (defined in Section~\ref{sec:intro}), we arrive at the following formula for difficulty. 
\begin{equation}
\label{eq:difficulty}
D_i = c_i T.
\end{equation}

\para{Commitment Constraints.} Because the difficulty is derived directly from miner commitments (Eq.~\ref{eq:difficulty}), it is possible for an attacker to falsely raise the difficulty arbitrarily high at a cost limited to bond $b$. To prevent this attack, it is important to ensure that every miner's commitment is realistic given their past hash rate. This constraint has ramifications for both bootstrapping and fully bonded miners (those with $nb$ coins in the bond pool). For a fully bonded miner, we stipulate that she can change her commitment by no more than some multiple $\mu$ times her average commitment over the previous $n$ blocks (e.g., $\mu=2$). 
As a protection against Sybil attacks\cite{Douceur:2002} described later, we allow up to fraction $\gamma$ of the total commitment for block $k_i$ to come from bootstrapping miners  (i.e., miners who have deposited fewer than $nb$ coins into the bond pool). If the aggregate commitments exceed fraction $\gamma$, then they are cut proportionately down to the maximum. Fraction $\gamma$, which is also tunable, should be fairly small (e.g., $\gamma=0.05$) because new miners have much less at stake than established miners (fewer coins in deposit), and the network has observed fewer blocks from which to assess their hash rate potential. The values for both $\mu$ and $\gamma$ should be set by the community at large (not just miners) and can also be updated regularly to respond to changes in miner composition.

\para{Mining Pools.} The Bonded Mining protocol is agnostic to the presence of mining pools, but the protocol does not greatly impinge on the ability to operate a pool effectively. Commitments must be made at the pool level, thus they can be changed only as often as the pool mines a block. A pool could aggregate constituent miner preferences in between blocks and update the overall commitment based on those preferences for the next block. Thus, if a pool mines a block approximately once an hour, then miners would be free to adjust their hash rate allocation to that pool with the same frequency.

\section{Threat Model}
\label{sec:attacks}
\label{sec:sybil}

\para{Primary Threats.} In the next sections, we focus our security analysis on attacks from miners who report their hash rate dishonestly. Rational miners intent on deviating from their commitment for $x$ blocks will report dishonestly when their \emph{preference} to deviate exceeds the greater of the value of \emph{coins at risk} or penalty associated with honestly reporting the deviations (i.e., $x(b - f^m)$). (We discuss coins at risk in Section~\ref{sec:detection} and preference in Section~\ref{sec:block_time}.) In this sense, Bonded Mining can ensure miners honor their commitment only to the extent that the fiat value of bond exceeds the value of mining on a different blockchain. There are numerous ways that an attacker can falsify reports, but in this preliminary work we focus on \emph{short-range} and \emph{long-range} attacks. A short-range attack involves an attacker significantly deviating from his commitment for a relatively small number of sequential blocks. In this context, we are primarily concerned with the attacker committing to a large hash rate, and subsequently dramatically lowering that hash rate, which would tend to increase block times during the attack. In contrast, the long-range  attacker deviates subtly from his commitment over the course of many blocks. Although not catastrophic during a short period of time, this attack can lead to systematic deviations from the target block time.

\para{Sybil Attacks.}
A single miner (or a cooperating coalition of miners) can always split his hash rate to appear to be multiple, lower hash rate miners in a Sybil attack~\cite{Douceur:2002}. Therefore, it is important to make incentives equitable for miners of any hash rate in order to encourage honest representation of affiliation and avoid the formation of mining cabals. We tune the Bonded Mining protocol to ensure that: \1 after Bootstrap, the amount of bond locked up at any given time is proportional to the miner's committed hash rate; and \2 the expected lockup time for bond and the expected probability of losing that bond are equal across hash rates for honest miners. We do not, however, attempt to align penalties for attacking miners with varying hash rates. That is to say, it will be possible that an attacker will suffer a lesser penalty by breaking up his hash rate among multiple identities. Note that this sort of asymmetry exists with other attacks. For example, a selfish mining attack only becomes possible for miners with a certain minimum percentage of the total hash rate.

\para{Other Attacks.}
 Bonded Mining's commitment validation test uses the inter-block times of each miner's {\em own} blocks; therefore, one miner's hash rate cannot influence the test results of another miner. We do not consider out-of-band denial-of-service attacks. A selfish mining (SM) attack~\cite{Eyal:2014} by one miner would alter inter-block times of the other miners by increasing their orphan rate. However, bonded miners could easily adjust their commitment to account for orphaned blocks due to an SM attack. Alternatively, Eq.~\ref{eq:difficulty} could be modified to take into account a miner's orphaned blocks~\cite{Grunspan:2019}. In fact, Bonded Mining makes it possible to attribute SM behavior to one or more miners who have a much lower orphan rate than other miners; we leave this analysis to future work. 
Doublespend (DS) attacks~\cite{Nakamoto:2009} are more challenging to carry out in Bonded Mining since the attacker must both succeed and avoid a commitment validation test failure in order to prevent loss of bond. Lastly, a miner could falsely report the timestamp in her own block headers. We do not investigate the impact of timestamp manipulation in this preliminary work, but note that the Bonded Mining protocol could impose restrictions on timestamps to greatly reduce the impact of such an attack. For example, it could stipulate that miners should synchronize clocks using NTP and that a miner should discard any block header that reports a timestamp deviating from their own clock by more than a reasonable block header propagation delay, perhaps 30 seconds.

\section{Report Validity Test} 
\label{sec:test}

We require a statistical test of the validity of a sequence of reports $r_1^m, \ldots, r_n^m$ from miner $m$. The test should have have both high precision and recall in order to simultaneously prevent malicious attacks while refraining from bond slashing honest miners with high probability. It should also be capable of simultaneously detecting various types of attacks. In this section, we describe a statistical test that effectively detects both short- and long-range attacks. It requires a sample of $n$ of the miner's most recent reports, where $n$ varies with her hash rate, but can otherwise be treated as a black-box test within the Bonded Mining protocol. 

Our approach is to use the popular one-sample Kolmogorov-Smirnov (KS) test~\cite{Massey-Jr.:1951} as a building block for test $\texttt{Valid}$. It is tempting to use a simpler point estimate of hash rate expected value as opposed to a goodness-of-fit (GoF) test over an entire distribution (which is provided by the KS test). There are two reasons that a distribution test is more desirable. First, the KS test is much more sensitive to systematic deviations in hash rate. Second, point estimates allow for attacks in which miners game the system by, for example, front loading all of their hash rate at the beginning of a test window; in that case the correct mean will be achieved, but the distribution will be wrong. 

Consider the random sequence $\mathbf{T}^m = T_1^m, \ldots, T_n^m$, denoting the inter-arrival times of blocks $k^m_1, \ldots, k^m_n$ mined by $m$. By $h_1^m, \ldots, h_n^m$ we denote the actual average hash rate for miner $m$, with $h_i^m$ associated with the time between blocks $k^m_{i-1}$ and $k^m_i$ ($h_i^m = r_i^m$ when $m$ reports hash rate honestly). Eq.~\ref{eq:difficulty} can be rewritten as 
\vspace{-1em}
\begin{equation}
\label{eq:ex_blk_tm}
E[T_i^m] = \hat{D}_i^m / h_i^m,
\end{equation}
where $\hat{D}_i^m$ is the average difficulty between blocks $k^m_{i-1}$ and $k^m_i$; i.e. the expected time required for $m$ to mine her $i$th block is equal to the average difficulty divided by her hash rate. 
It is well known~\cite{Rizun:2016,Bissias:2017} that each $T_i^m$ follows the exponential distribution. That is,  $T_i^m \sim \texttt{Expon}(\hat{D}_i^m / h_i^m)$.
And because the distribution is a scale family, it is straightforward to show that $(T_i^m h_i^m / \hat{D}_i^m ) \sim \texttt{Expon}(1)$. Now define statistic $\mathbf{X}^m = X^m_1, \ldots, X^m_n$ such that $X^m_i = T_i^m r_i^m / \hat{D}_i^m$. When $m$ reports honestly, we have \mbox{$X_i^m \sim \texttt{Expon}(1)$}.

\begin{figure}[t]
\hspace{-10pt}\begin{minipage}[c]{0.575\textwidth}
\includegraphics[trim=0 0 0 30,clip,width=\linewidth]{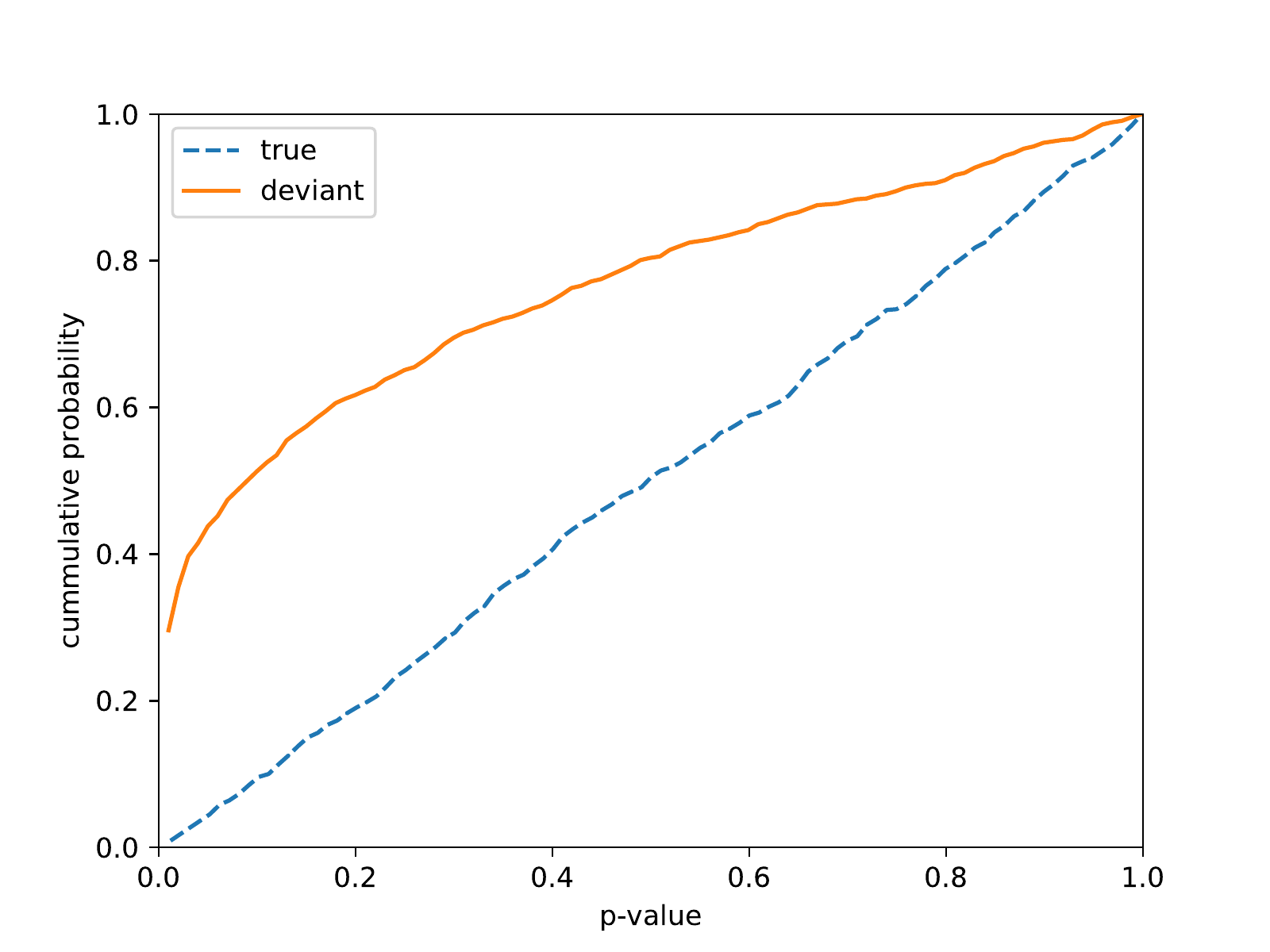}
\end{minipage}
\begin{minipage}[c]{0.45\textwidth}
\caption{Empirical cumulative distribution of $p$-values over 1000 trials for 500 samples drawn from  the \emph{true} distribution $\texttt{Expon}(1)$ (blue, dashed), and a \emph{deviant} sequence of exponential random variables with mean drawn from a random walk about $1$ with standard deviation equal to $0.01$ (orange, solid). The true samples manifest $p$-values according to Eq.~\ref{eq:pvalue}, in that for any given $x \in [0,1]$, approximately $x$ $p$-values fall under value $x$.}
\label{fig:pvalues}
\end{minipage}\vspace{-2ex}
\end{figure}

The one-sample Kolmogorov-Smirnov (KS) test is a statistical test of the null hypothesis $\mathcal{H}_0$ that samples $\mathbf{x}^m = x_1^m, \ldots, x_n^m$, collectively an instance of $\mathbf{X}^m$, were drawn from a given distribution. In our case, the distribution is $\texttt{Expon}(1)$. The KS statistic is given by $\Delta_n = \sup_x |S_n(x) - F(x)|$, where $F(x)$ is the cumulative distribution function for $\texttt{Expon}(1)$ and $S_n(x)$ is the empirical distribution function derived from samples $\mathbf{x}^m$. The alternative hypothesis $\mathcal{H}_1$ contends that the samples were not drawn from $\texttt{Expon}(1)$. Now let $p(\Delta_n)$ be the $p$-value for $\Delta_n$ and define $\delta_i$ to be a realization of $\Delta_n$. By definition~\cite{Casella:2002}, 
\begin{equation}
\label{eq:pvalue}
P(p(\Delta_n) < \tau ~|~ \mathbf{X}^m \sim \texttt{Expon}(1)) < \tau.
\end{equation}
Therefore, assuming that $\mathbf{x}^m$ was drawn from $\texttt{Expon}(1)$, we are guaranteed that the rejection region $p(\delta_n) < \tau$ will carry probability of type-I error (falsely rejecting $\mathcal{H}_0$) no greater than $\tau$. Accordingly, large $p$-values provide evidence in support of $\mathcal{H}_0$ and low values provide evidence in support of $\mathcal{H}_1$. 

Figure~\ref{fig:pvalues} shows the empirical cumulative distribution for $p(\delta_n)$ drawn from two different distributions: \emph{true} and \emph{deviant}.  The true samples (blue, dashed curve) are drawn from distribution $\texttt{Expon}(1)$ while the deviant samples (orange, solid curve) are drawn from a sequence of exponential random variables with mean changing according to a random walk about $1$ with standard deviation equal to $0.01$. The true samples produce nearly perfect $p$-values in that approximately fraction $x$ of all trials fall below $p$-value $x$, indicating an accurate estimation of type-I error. The deviant samples manifest significantly lower $p$-values, indicating high confidence in $\mathcal{H}_1$. Approximately 30\% of deviant samples register a $p$-value very close to 0. We summarize a binary test based on the KS statistic below.

\vspace{-3pt}
\begin{mydef}
For tunable threshold $\tau \in [0,1]$, test $\texttt{KS}(\mathbf{r}^m; n, \tau)$ is equal to 1 when the $p$-value $p(\delta_n)$ of KS statistic $\Delta_n$ exceeds $\tau$ and 0 otherwise. Accordingly, the probability of a type-I error is equal to $\tau$.
\end{mydef}

Because we are interested in detecting both short- and long-range attacks, we apply two KS tests: one over a short window $n_s$; and the other over a longer window $n_l$. Associated with these tests are the thresholds $\tau_s$ and $\tau_l$, respectively. The window sizes and thresholds vary with both test type and the miner's committed hash rate. We combine the two tests into one to form $\texttt{Valid}$ by multiplying their output. The probability of a type-I error in $\texttt{Valid}$ is bounded by $\tau_s + \tau_l$ because it occurs if there is a type-I error in either of the KS tests.

\vspace{-3pt}
\begin{mydef}
$\texttt{Valid}(\mathbf{r}^m; n_s, n_l, \tau_s, \tau_l)$ is equal to 1 when both $\texttt{KS}(\mathbf{r}^m; n_s, \tau_s)$ and $\texttt{KS}(\mathbf{r}^m; n_l, \tau_l)$ are equal to 1, and it is 0 otherwise. The probability of a type-I error is bounded by $\tau_s + \tau_l$.
\end{mydef}

\section{Attack Risk and Detection}
\label{sec:detection}

In this section, we quantify the \emph{coins at risk} for miners who report their hash rate dishonestly, which we regard as an attack. When a miner fails test $\texttt{Valid}$, her entire bond of $n b$ coins  are lost and she must proceed through the Bootstrapping state  to resume mining. Thus, at a minimum, the coins at risk are the expected amount of bond lost when test $\texttt{Valid}$ fails. But, any failure is also a significant step back for a miner, the $\gamma$ parameter (see Section~\ref{sec:protocol}) limits her committed hash rate during bootstrap (a period of about 70 days), which amounts to an opportunity cost in terms of coinbase revenue. 

We ran a  simulator of PoW block mining to test the ability for a dishonest miner, with varying fraction of the total hash power, to conceal short- and long-range deviations from her commitment. We evaluated miners in three categories: \1 \emph{honest}, \2 \emph{short-range dishonest}, and \3 \emph{long-range dishonest}.  All miners varied their commitment each time they generated a block by performing a random walk with standard deviation equal to 1\% of their originally committed hash rate. Honest miners always reported this deviation, while long-range dishonest miners reported that their commitment never changed from its original value. Short-range dishonest miners were honest about their long-range deviations in hash rate (just like honest miners), but at the end of each test window they mined at 1/5 of their committed hash rate for approximately one week, while reporting no deviation from their commitment.

We generated block creation times by sampling from $\texttt{Expon}(\alpha T)$, where $\alpha$ was given by the ratio of total hash rate $h_i$ to actual miner hash rate $h_i^m$. ($\alpha T$ is equivalent to Eq.~\ref{eq:ex_blk_tm} when actual hash rates are known.) Each block was randomly assigned to either the honest or dishonest groups with a probability proportional to the fraction of total hash rate possessed by the dishonest group. 

\para{Test Application.} There are two separate, but related, goals for test $\texttt{Valid}$. First, given a single sequence of $n$ consecutive blocks called a \emph{test window}, we require that $\texttt{Valid}$ can distinguish between honest and either short- or long-range dishonest miners by means of a statistical test on  sub-windows of size $n_s$ and $n_l$ within the those $n$ blocks. Second,  to understand the probability of ongoing attack success, we also require that $\texttt{Valid}$ continue to differentiate between honest and either short- or long-range dishonest miners during a long temporal sequence of overlapping test windows. The two goals are met by applying test $\texttt{Valid}(\mathbf{r}^m; n_s, n_l, \tau_s, \tau_l)$ with the long window $n_l = n$, short window $n_s \ll n$, short threshold $\tau_s$, and long threshold $\tau_l$ all defined so that the probability of a type-I error remains extremely low for an entire year of mining. Our metrics for success are type-I error and attack detection rate (i.e., the rate at which $\texttt{Valid} = 0$ when applied to the attacker's test windows). The former should be sufficiently close to 0 so as to encourage honest miners. And the latter should be somewhat higher than 0 so as to discourage dishonest miners. 
These parameters can be chosen offline and independent of blockchain conditions.

\para{Parameter Selection.} Table~\ref{tbl:params} shows the parameter choices we made, which we applied to all experiments reported below. In general, lower hash rate miners (as a percentage of the total) require smaller windows for $\texttt{Valid}$ to produce dishonest-detection rates that would be significant to attackers. This is a desirable property because lower hash rate miners require more time to produce blocks. We chose $n_s$ and $n_l$ so that the expected temporal duration of each test sub-window was the same for miners across all hash rates (approximately 1.5 days for $n_s$ and 70 days for $n_l$). We then chose the values for $\tau_s$ and $\tau_l$  to render the probability of type-I error close to 0, even after one year of expected mining time. These parameter choices are validated in the next sections. 

\begin{table}[t]
\centering
\caption{Parameters choices for test $\texttt{Valid}$ given committed hash rate as a percentage of the total. Short- and long-range test sub-window sizes are reported as $n_s$ and $n_l$, respectively, while short- and long-range tolerances are reported as $\tau_s$ and $\tau_l$, respectively.\\}
{\relsize{-1}
\begin{tabular}{rrr rrr rrr rrr rrr}
\toprule
\multicolumn{3}{c}{\bf committed }&
\multicolumn{3}{c}{\bf short} &
\multicolumn{3}{c}{\bf long}  & 
\multicolumn{3}{c}{\bf short} &
\multicolumn{3}{c}{\bf long}  \\
\multicolumn{3}{c}{\bf hash rate}&
\multicolumn{3}{c}{\bf \mbox{window $n_s$}} &
\multicolumn{3}{c}{\bf \mbox{window $n_l$}}  & 
\multicolumn{3}{c}{\bf threshold $\tau_s$} &
\multicolumn{3}{c}{\bf \mbox{threshold $\tau_l$}}  \\
\cmidrule(lr){1-3}\cmidrule(lr){4-6}\cmidrule(lr){7-9}\cmidrule(lr){10-12}\cmidrule(lr){13-15}
&1\%  &&&2   &&& 100  &&& $10^{-7\phantom{1}}$  &&&$10^{-7\phantom{1}}$  \\
&10\% &&& 20 &&& 1000 &&& $10^{-10}$ &&& $10^{-10}$\\
&25\% &&& 50 &&& 2500 &&& $10^{-12}$ &&& $10^{-12}$  \\
&50\% &&&100 &&& 5000 &&& $10^{-12}$ &&&  $10^{-12}$  \\
\bottomrule
\end{tabular}
}
\label{tbl:params}
\end{table}

\para{Attack Detection.} In Table~\ref{tbl:params} we specify short- and long-range test sub-windows for committed hash rates varying from 50\% down to 1\% of the network total. Given these parameters, the short-range sub-window for a miner committed to 1\% of the hash rate is $n_s = 2$.  To discourage Sybil attacks  parameters must be chosen equitably for honest miners with any hash rate, as we argue in Section~\ref{sec:sybil}. This implies that, in its current incarnation, Bonded Mining cannot support miner commitments less than 0.5\% of the total hash rate, because this would require a value for $n_s$ that is less than 1. Note that this limitation is tied to the amount of proof of work we receive from miners; i.e., we cannot detect an attack during periods of time when the miner does not produce blocks. 
For BTC and BCH, respectively, 99.4\% and 98.6\% of the total hash rate from March 24--June 24, 2019
was contributed by miners with at least 1\% of the hash rate.\!\footnote{See \url{https://btc.com/stats/pool} and  \url{https://bch.btc.com/stats/pool}.} Smaller miners do not need to be excluded from the system as they can join a larger pool.  
And in future work, we will explore  PoW schemes such as Bobtail~\cite{Bissias:2017} and FruitChains~\cite{Pass:2017} that have  miners  broadcast additional PoW information. For example, Bobtail could easily decrease Bonded Mining's minimum allowable hash rate by several orders of magnitude.

\subsection{Accuracy of $\texttt{Valid}$ Over the Bootstrapping Window}

\begin{table}[t]
\centering
\caption{\textbf{[Attacks During Bootstrapping]} Results over 1000 trials of test $\texttt{Valid}$ for various attacker hash rates (as a percentage of the total) and parameters selected according to Table~\ref{tbl:params}. All results in this table correspond to dishonest miners who deviate from their committed hash rate during bootstrap. Long-range attackers deviate  according to a random walk with standard deviation equal to 1\% of their commitment. Short-range attackers drop to 1/5 of their committed hash rate for the last $n_s$ blocks in the  window, where $n_s$ is determined by Table~\ref{tbl:params}. See Fig.~\ref{fig:short_long} for rates after bootstrapping.\\}
{\relsize{-1}
\begin{tabular}{crcrc}
\toprule
\bf Committed  && \bf Short-range && \bf Long-range  \\
\bf hash rate && \bf  detection rate && \bf detection rate  \\
\cmidrule{1--2}\cmidrule{3--4}\cmidrule{5--6}
~1\% && 0.033 && 0.271  \\
10\% && 0.108 && 0.653 \\
25\% && 0.837 && 0.660 \\
50\% && 1.000 && 0.546 \\
\bottomrule
\end{tabular}
}
\label{tbl:ks_window}
\end{table}

Table~\ref{tbl:ks_window} shows the probability of attack detection at the end of the bootstrapping window (block $n$) for   short- and long-range attackers. Results are based on parameters from Table~\ref{tbl:params} for four different hash rate percentages, averaged over 1000 trials for each.
We do not show the rate of type-I error when miners are honest because not a single error was encountered in the 1000 trials.
However, by construction, the probability of a type-I error in any given test window is bounded by $\tau_s + \tau_l$.

The most important detection rate is for the 1\% miner because a high hash rate miner  can  masquerade as multiple low hash rate miners. Thus,  a determined attacker can avoid detection during the Bootstrapping state with 96.7\% probability. From Table~\ref{tbl:params}, $n=n_l=100$; and thus there are $0.03 n b = 3b$ coins at risk for the attacking miner. We show next that any initial success will likely be short-lived and the coins at risk will rise if the attacker continues as a fully bonded miner. 

\subsection{Attacks by Fully Bonded Miners}
\label{sec:attack_fully_bonded}

Once in the Fully Bonded state, a miner can remain in that state by continuing to add a new commitment for each additional block mined. $\texttt{Valid}$ is tested against the sliding window of the $n$ most-recent blocks generated by the miner. These sliding windows are highly dependent since adjacent windows share all but one block. If any single test fails, the total bond of $nb$ is lost.  
  
\begin{figure}[t]
\hspace{-1.5em}{\begin{minipage}[c]{0.55\textwidth}
\includegraphics[trim=0 0 20 20,clip,width=1.1\linewidth]{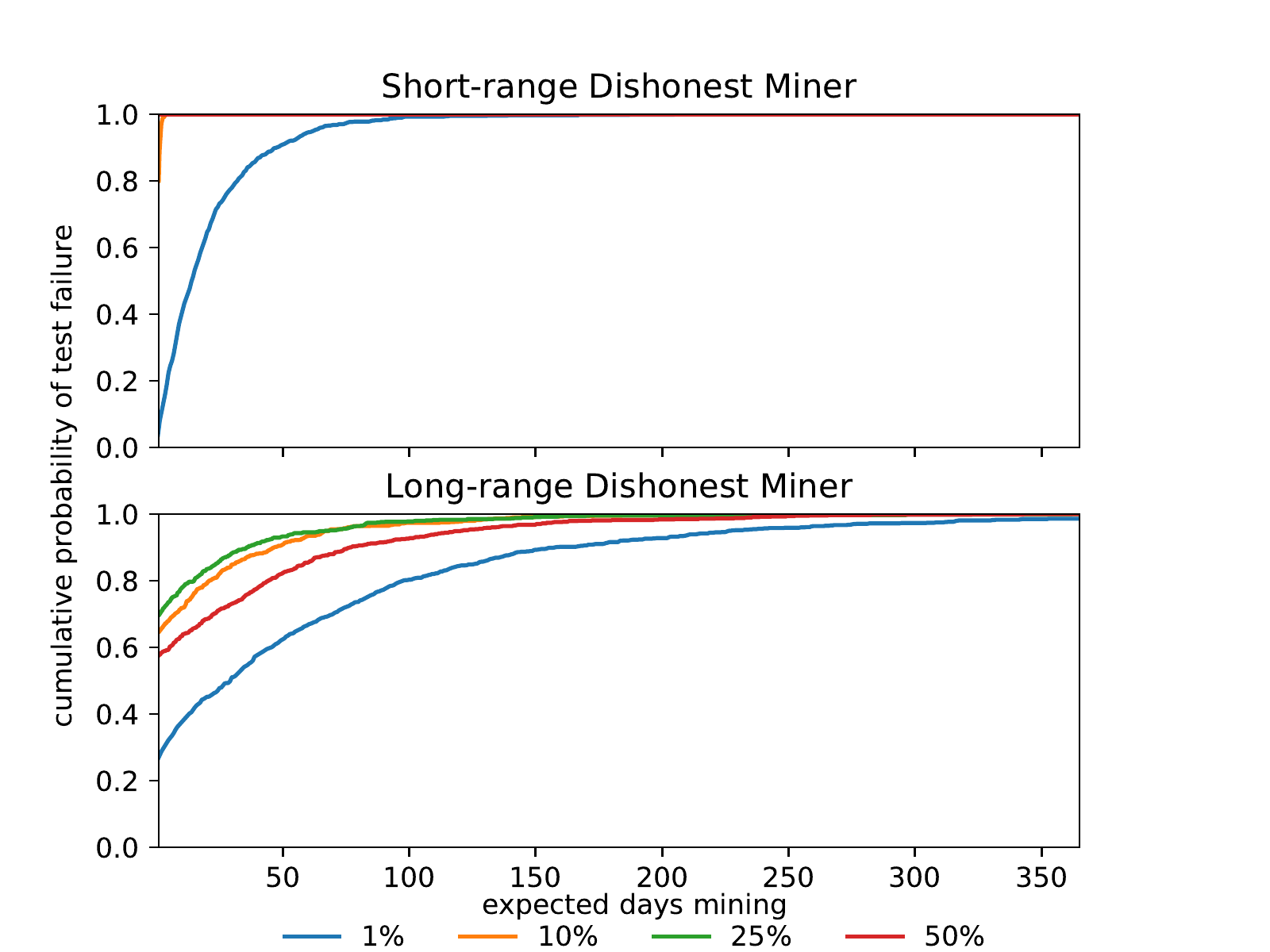}
\end{minipage}
\begin{minipage}[c]{0.49\textwidth}
\caption{\textbf{[Attacks by fully bonded miners.]} Probability that at least one test window fails $\texttt{Valid}$ for short-range (top) and long-range (bottom) attacks spanning up to one year in duration. Short-range curves not visible are extremely close to probability 1.0 at nearly all times. (Note, for those not reading in color: in the top plot, the 50\% curve appears above the 1\% curve; and in the bottom plot, curves appear in the following order from top to bottom: 25\%, 10\%, 50\%, 1\%.)}
\label{fig:short_long}
\end{minipage}}\vspace{-2ex}
\end{figure}

We simulated one year's worth of blocks from attacking miners subject to Table~\ref{tbl:params} parameters, and we determined the mean probability (across 1000 trials)  that \texttt{Valid} detects at least one attack during the year.
Results in Figure~\ref{fig:short_long} show that approximately 30 days after bootstrapping, even attacks by 1\% miners are detected with 50\% probability. Thus,   since again $n=n_l=100$,  the coins at risk after 30 days of attack is $0.5 n b = 50 b$.

\para{Probability of type-I error}. We ran the same experiment for honest miners who similarly varied their hash rate as in the short- and long-range attacks for multiple test windows spanning a year in duration, except that in these experiments miners reported their hash rate honestly. Over the course 1000 trials for each set of parameters selected from Table~\ref{tbl:params}, test $\texttt{Valid}$ never returned a false positive in any test window throughout the simulated year of mining across all hash rate commitments and for both short- and long-range deviations. 
Therefore, the  probability of type-I error is no higher than 0.003 based on a   
95\% confidence interval from the 1000 trials~\cite{hanley:1983}.
Overall,  miners that honestly report their commitments are very unlikely to fail test $\texttt{Valid}$.

\section{Block Time Stability}
\label{sec:block_time}

Section~\ref{sec:test} introduces a hypothesis test for determining when a miner reports his hash rate honestly over a window of $n$ consecutive blocks that he has mined. If the miner fails this test, his bond of $nb$ coins is lost; and if he misbehaves such that his probability of failure is $p$, then we say that he has $pnb$ \emph{coins at risk}. On the other hand, an honest miner, who passes the test, is eligible to submit a transaction in the next block that reconciles the bond deposited for the first block in the window, with the reconciliation fee $f^m$ being paid according to Eq.~\ref{eq:reconcile}. Thus, when deviating from his commitment for $x$ blocks, a rational miner will report honestly when $pnb$ exceeds $x (b-f^m)$. Section~\ref{sec:detection} quantified $pnb$ for various attacks. In this section, we assume that coins at risk are high enough that the miner always reports his hash rate honestly (i.e. $pnb > x (b-f^m)$). Nevertheless, he might still change his commitment dramatically or even deviate significantly from his commitment (all considered honest behavior). Naturally, these deviations can affect the performance of the DAA. Thus, our goal is to \1 quantify the affect on block times given various magnitudes of hash rate deviation, and \2 quantify the financial cost to a miner who performs these deviations.

\begin{figure}[t]
\hspace{-.07\linewidth}\begin{subfigure}[h]{0.49\linewidth}
\includegraphics[trim=0 0 0 10,clip,width=1.1\linewidth]{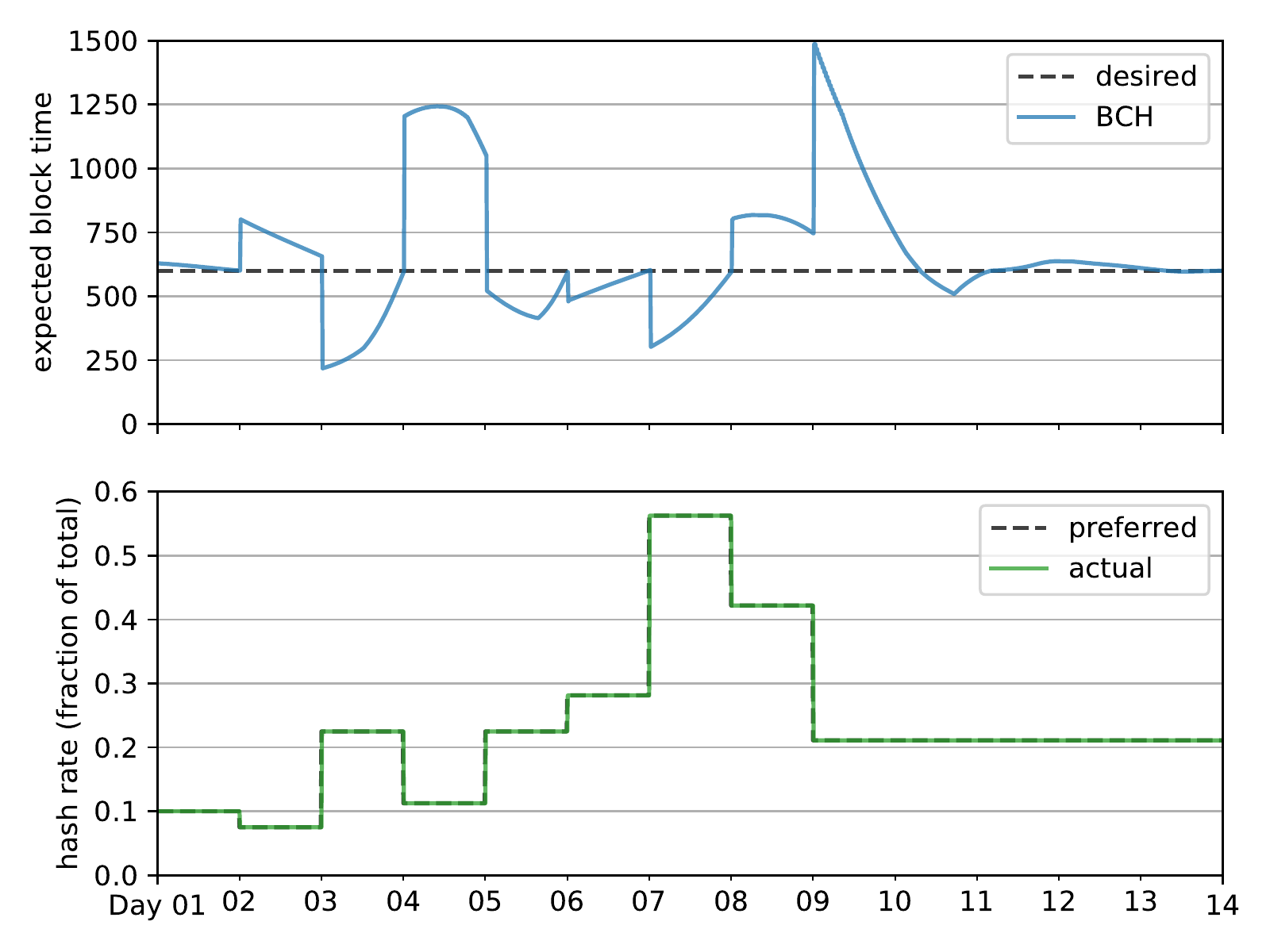}
\end{subfigure}
\hfill
\begin{subfigure}[h]{0.49\linewidth}
\includegraphics[trim=0 0 0 10,width=1.1\linewidth]{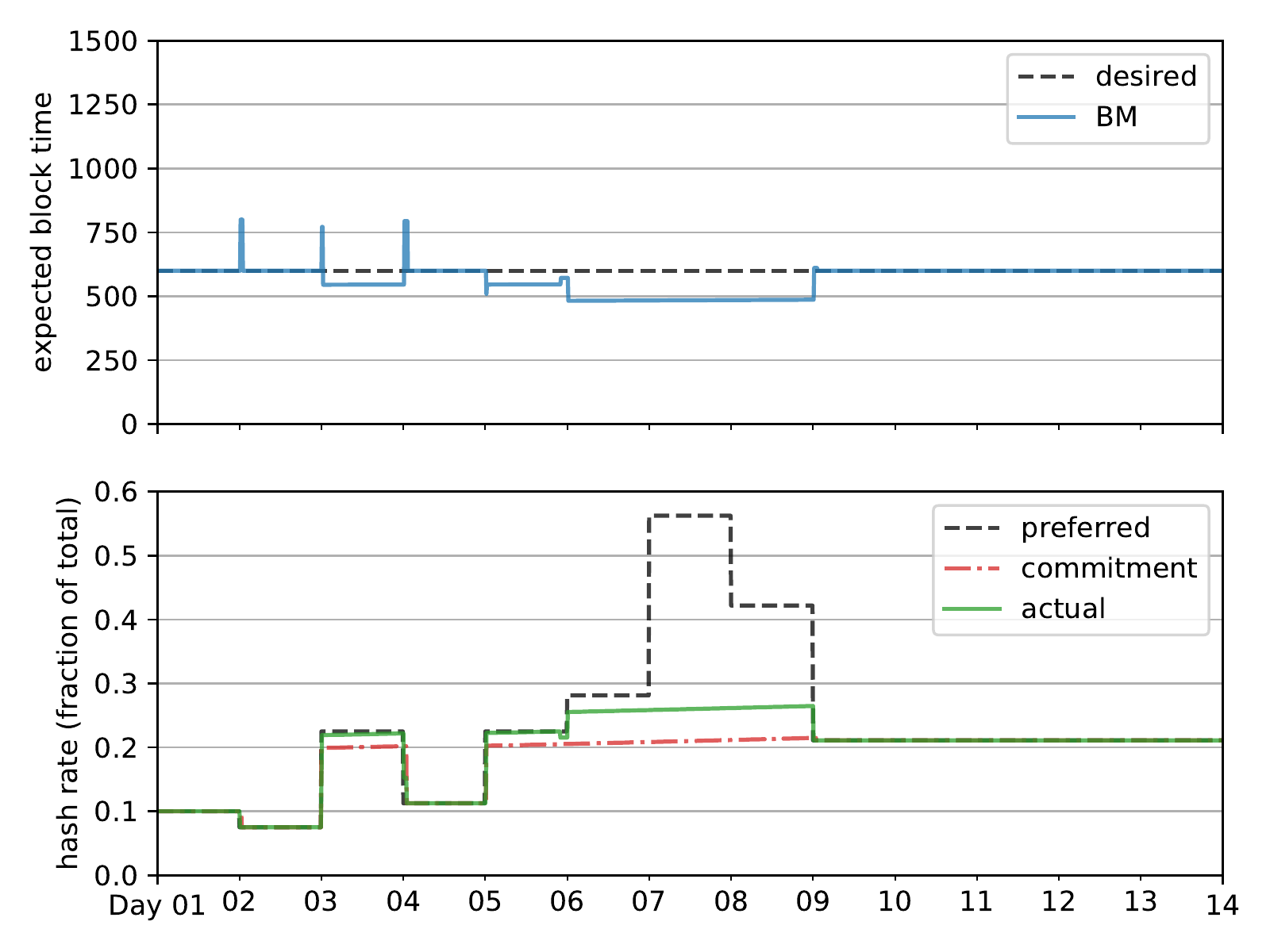}
\end{subfigure}%
\caption{Comparison of expected block times in simulation for (left) the Bitcoin Cash protocol (BCH) and (right) the Bonded Mining protocol (BM) where miner cost tolerance is $\kappa = 0.25$. Bottom facets show fluctuations in miner hash rates, while the top facets show the block time that results from the DAA response to those fluctuations. The target block time for both protocols is $T = 600$ seconds. Relative to BCH, The BM DAA deviates less from desired block time in terms of amplitude and duration.}
\label{fig:bch_vs_bonded}\vspace{-2ex}
\end{figure}

\subsection{Block Time Simulation} 

To achieve the goals of this section, we created an expected block time simulation that takes aggregate miner hash rate as input, generates blocks according to current hash rate and difficulty, and runs a 
DAA to update the difficulty. We compared two scenarios: one using the DAA for Bitcoin Cash (\emph{BCH}, for brevity); and the other using the Bonded Mining protocol DAA (\emph{BM}, for brevity). We chose to compare against  Bitcoin Cash because it uses one of the newest and most dynamic DAAs currently deployed. Because the simulation reports expected block time, it is deterministic given miner preferences (Table~\ref{tbl:hash_prefs}). The simulation is meant to highlight DAA behavior in a variety circumstances, we leave more systematic analysis to future work. 

\begin{table}[h]
\centering
\caption{Hash rate preferences, expressed as a percentage of total available hash rate, for miners during two weeks of simulated mining. The stated preference extends from the stated day until the next stated day or the end of the experiment.\\}
{\relsize{-1}
\begin{tabular}{ccccccccc}
\toprule
Day 1 & Day 2 & Day 3 & Day 4 & Day 5 & Day 6 & Day 7 & Day 8 & Day 9  \\
\cmidrule(lr){1-1}\cmidrule(lr){2-2}\cmidrule(lr){3-3}\cmidrule(lr){4-4}\cmidrule(lr){5-5}\cmidrule(lr){6-6}\cmidrule(lr){7-7}\cmidrule(lr){8-8}\cmidrule(lr){9-9}
10\% & 7.5\% & 22.5\% & 11.3\% & 22.5\% & 28.1\% & 56.3\% & 42.2\% & 21.1\% \\
\bottomrule
\end{tabular}
}
\label{tbl:hash_prefs}
\end{table} 

\para{Blocks.} Both BCH and BM targeted block time $T = 600$ seconds. In both cases, the expected block time was calculated as $T\cdot D_i / h_i$ (a relationship that follows from Eq.~\ref{eq:fundamental}), where $h_i$ and $D_i$ are the total hash rate and difficulty, respectively, during the period when block $i$ is mined.

\para{Preference.} At any given time we assumed that miners had a \emph{preference} for what percentage of their hash rate they would like to apply to mining on the Bonded Mining blockchain. The notion of preference captures a miner's tendency to either divert her hash rate to another blockchain or stop mining altogether. Preferences could change, for example, because of changes in coin or energy prices. We do not  model miner preferences but rather treat them as input. 

\para{Hash rate.}  For simplicity, we adjusted all hash rates in unison every block (as though all miners acted in concert). But we set the block window to $n = 1000$ and updated commitments only every 10th block as if the network was comprised of 10 miners, each having 10\% of the total hash rate. Over the course of two weeks of simulated mining time, miners developed the preference for a given hash rate at different times, expressed as a percentage of their total hash rate available (see Table~\ref{tbl:hash_prefs}). For BCH, miners immediately realized their hash rate preference. However, for BM, miners were restricted in two ways: \1 we set the commitment constraint to  $\mu=2$ (see Section~\ref{sec:protocol}) so that miners could not change their commitment by more than twice the average of their $n$ earlier commitments; and \2 miners were averse to deviating from their commitments because such deviations would result in a loss of bond. Note that BM miners were assumed to be fully bonded, so parameter $\gamma$ was not relevant.

\para{Cost tolerance.} Miners obeyed a variable \emph{cost tolerance}, $\kappa$, expressed as a fraction of per-block bond $b$, which is the amount of bond that they were willing to lose per block due to deviation from their commitment. For example, if $\kappa = 0.1$ and $b = 100$ USD, miners are willing to lose up to $10$ USD per block by deviating from their commitment if there exists a 10 USD opportunity cost for mining on the present chain over another. 

\para{Difficulty adjustment.} The target block time for the simulation was $T = 600$ seconds. BCH used the Bitcoin Cash DAA~\cite{Sechet:2017}: the rolling sum (over the last 144 blocks) of difficulty $D$ (which gives the expected number of hashes performed) and block time $M$ are calculated; the block time sum is clamped with a high and low pass filter: $M' = \pmb{\max}\{72 \cdot T, \pmb{\min}\{M, 288 \cdot T\}\}$; and the new difficulty becomes $D/M'$. The DAA for BM was Eq.~\ref{eq:difficulty}: $D_i = c_i T$, which is the product of target block time and total committed hash rate.

\subsection{Simulation Results}

\begin{figure}[t]
\begin{minipage}[c]{0.63\textwidth}
\includegraphics[trim=0 0 0 95,clip,width=\linewidth]{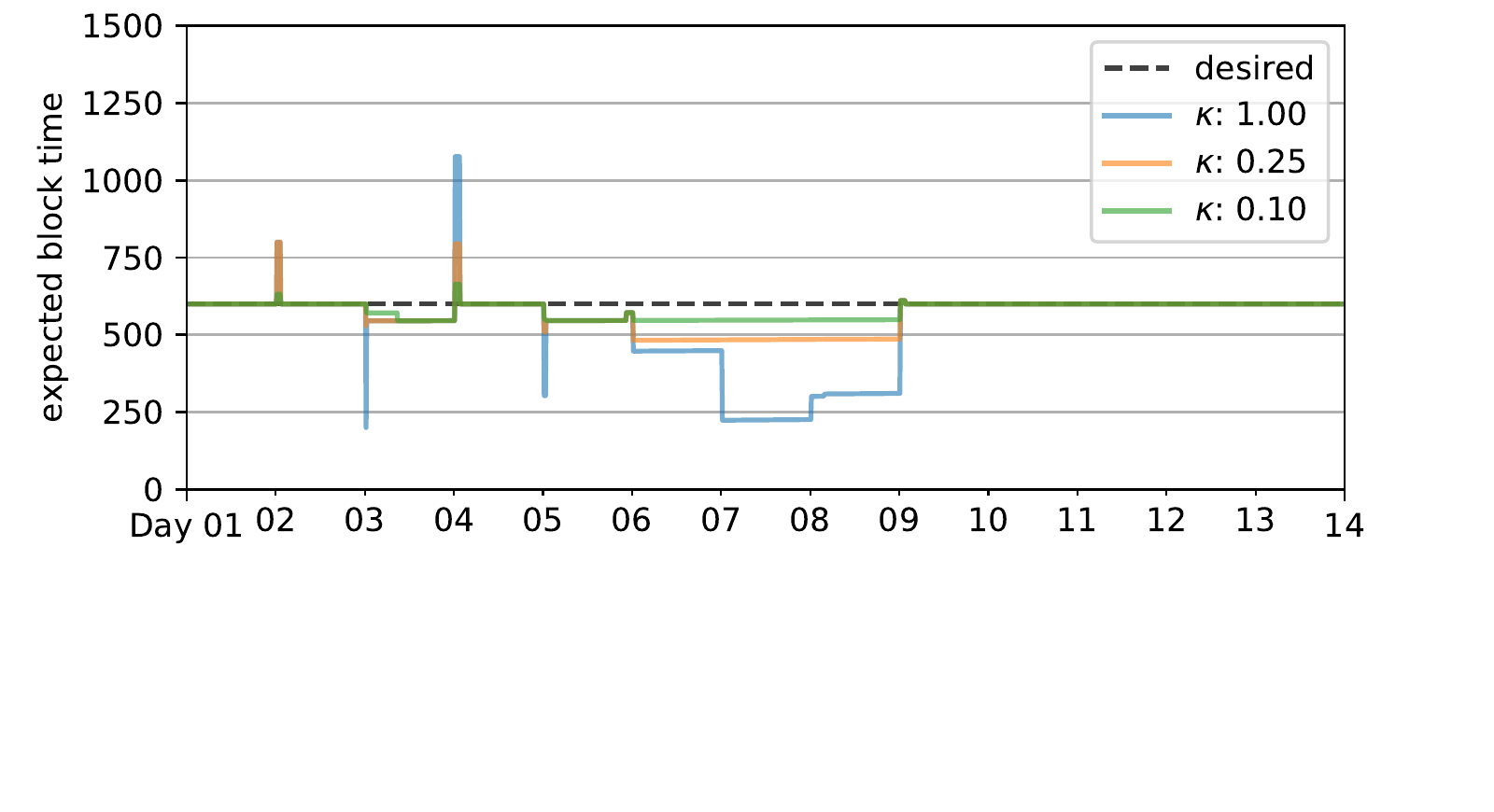}
\end{minipage}
\begin{minipage}[c]{0.35\textwidth}
\caption{Comparison of expected block times using Bonded Mining with varying cost tolerances $\kappa$. The target block time was $T = 600$ seconds. (Note, for those not reading in color, $\kappa=1.0$ is the curve that varies most from desired, followed by $\kappa = 0.25$, and $\kappa = 0.1$, which has the lowest variance from desired.)
}
\label{fig:multiple_tolerances}
\end{minipage}\vspace{-6ex}
\end{figure}

During the two weeks of simulated mining time, hash rate preferences changed according to Table~\ref{tbl:hash_prefs}. Figure~\ref{fig:bch_vs_bonded} compares the resulting expected block times for each. BCH (left plot) shows several interesting features. First, we can see that expected block time deviates significantly from the target time during periods immediately following a major change in hash rate: the most dramatic drop is to 250 seconds, and the most dramatic rise is to nearly 1500 seconds. Moreover, these deviations were not corrected by the DAA for up to one day following the change in hash rate. Second, even during times when hash rate remains unchanged (for examples the period after Day 11, the expected block time oscillates about the target time. We suspect that this phenomenon is due to a feedback cycle that emerges from the DAA, which itself is essentially a feedback controller.

In contrast, Figure~\ref{fig:bch_vs_bonded} (right) shows that BM has much lower amplitude and duration of deviation from expected block time relative to BCH. And when hash rate is not changing, BM is able to maintain an expected block time that exactly matches the target time.

The Bonded Mining protocol is able to keep expected block time close to the target time by controlling the change in commitment and the tendency to deviate from that commitment. Miners are not allowed to vary their commitments by more than factor $\mu = 2$ from the average of previous commitments. Nevertheless, they may still vary their \emph{actual} hash rate as long as they can tolerate the penalty. Figure~\ref{fig:bch_vs_bonded} (right) shows that miners with cost tolerance $\kappa = 0.25$ will follow their hash rate preference, but only up to a point. When the preference deviates far beyond what the factor $\mu$ allows, miners cease to change their hash rate in order to avoid bond loss beyond 25\% per block. Figure~\ref{fig:multiple_tolerances} shows how the expected block time begins to deviate more and more as the cost tolerance $\kappa$ increases up to 1.0. What is not shown in the figure is that for $\kappa = 1.0$, the hash rate preference is followed perfectly, meaning that miners sacrifice as much bond as is necessary to meet their preference. For $\kappa=1.0$, the worst-case scenario for Bonded Mining, Figure~\ref{fig:multiple_tolerances} also shows that while the duration of deviation from the target block time is longer than for BCH, the maximum deviation is less. 

\subsection{Setting the Bond}

We have intentionally not specified how the community should choose the per-block bond value $b$. It is a security parameter that trades off between accessibility and security of mining, and the correct tradeoff is subjective. For example, consider  setting $b$ equal to 1/10 the coinbase value. This requires a fully bonded miner (with any committed hash rate) to lockup the equivalent of approximately 10\% of her mining profit for 70 days. This financial commitment is a strong deterrent for would-be attackers since failing even a single test $\texttt{Valid}$ results in a total loss of bond. We do note that it could be a significant hardship for miners who operate on slim profit margins. 

\section{Related Work}

Considerable effort has been devoted to refining the PoW difficulty adjustment algorithm (DAA) in the context of  feedback control. In comparison to Bonded Mining, these past approaches do not solicit information on future hash rate from miners. They suffer from oscillations and delay in reaching the target block time because they are reactive rather than proactive.

Alternative DAAs include methods by Kraft~\cite{Kraft:2016},  Meshkov et al.~\cite{Meshkov:2017}, Fullmer and Morse~\cite{Fullmer:2018}, and Hovland et al.~\cite{Hovland:2017}. Meshkov et al.\ note that alternating between coins has an effect on inter-block time.  Grunspan and P\'erez-Marco~\cite{Grunspan:2019} recently proposed a modified DAA that includes uncle blocks in the difficulty calculation to thwart selfish mining. 
The recent fork of Bitcoin Cash (BCH)~\cite{bitcoin-cash} from the core Bitcoin blockchain~\cite{bitcoin-core} (BTC) sparked a flurry of DAA research.  
The principal concern was 
to mitigate affects of \emph{fickle mining}~\cite{Kwon:2019}, where miners abruptly move their hash power from the BCH to the BTC blockchain (or vice versa).
See also Aggarwal and Tan for an analysis of mining during that time~\cite{Aggarwal:2019}.
S\'echet devised the algorithm {\em cw-144}~\cite{Sechet:2017}, which is currently used in BCH; it uses the ratio of the rolling average of chain work to block time and was based on an early model from Kyuupichan~\cite{Booth:2017}.  
Harding's {\em wt-144}~\cite{Harding:2017} is a similar solution, weighting block time by recency and block target. Stone~\cite{Stone:2017b} proposed adding \emph{tail removal} to the algorithms above, which drives down the difficulty within a block interval in order to guarantee liveness and dampen block time oscillations.

We use a bond as collateral for a  pledge to perform an offered hash rate over a specific period of time. Similarly, Proof of Stake (PoS) protocols (including hybrid PoW/PoS)~\cite{Bentov:2016,Duong:2018,Buterin:2017} accept collateral as a pledge to act honestly in validating transactions over a specific period of time. We do not intend our work to be a replacement for PoS, but note that Bonded Mining similarly provides a fixed set of consensus participants that in our case are validated by PoW. For hybrids of PoW/PoS and hybrids of PoW/BFT\cite{Decker:2016,Kogias:2016}, Bonded Mining can bolster the performance of the PoW component. 

\section{Conclusion}

We presented Bonded Mining, a proactive DAA for PoW blockchains that sets the mining difficulty based on bonded commitments from miners. The protocol incentivizes a miner to honor his commitment by confiscating a fraction of bond that is proportional to the deviation from his commitment. We developed a statistical test that is capable of detecting short- and long-range deviations from commitments with very low probability of falsely implicating honest miners. In simulation, and under reasonable assumptions, we showed that Bonded Mining is more successful at maintaining a consistent inter-block time than is the DAA of Bitcoin Cash, one of the newest and most dynamic DAAs in production. The present work is preliminary, having the following limitations: \1 currently the lowest hash rate miner that it supports has 1\% of the total; \2 we have only demonstrated defenses against two types of attacks; and \3 we only directly compare Bonded Mining to the Bitcoin Cash DAA. Future work will seek to address these limitations. We recommend these limitations be fully investigated before deployment.

\urlstyle{sf}
\pagestyle{plain}

{\footnotesize \bibliographystyle{acm}
\bibliography{references}}

\appendix

\section{List of Symbols}

\centering
{\relsize{-1}
\begin{tabular}{r l}
\toprule
{\bf Symbol } &~~ {\bf Description} \\
\cmidrule(lr){1-2} \cmidrule(lr){1-2}
$m$ &~~ miner \\
$n, n_s, n_l$ &~~ blocks in test window: generic, short, long \\
$b$ &~~ bond, a portion of coin held until miner passes test $\texttt{Valid}$ \\
$k_i^m$ &~~ $i$th block mined by $m$ \\
$c_i$ &~~ committed hash rate for entire network between blocks $i-1$ and $i$ \\
$c_i^m$ &~~ hash rate commitment for $m$ between blocks $k_{i-1}^m$ and $k_i^m$ \\
$h_i$ &~~ actual hash rate for entire network between blocks $i-1$ and $i$ \\
$h_i^m$ &~~ actual hash rate for $m$ between blocks $k_{i-1}^m$ and $k_i^m$ \\
$r_i^m$ &~~ reported hash rate for $m$ between blocks $k_{i-1}^m$ and $k_i^m$ \\
$f_i^m$ &~~ reconciliation payment to $m$ in block $k_i^m$ \\
$T_i^m$ &~~ time between $k_{i-1}^m$ and $k_i^m$ \\
$T$ &~~ target inter-block time \\ 
$D_i$ &~~ mining difficulty (expected number of hashes per block) between blocks $i-1$ and $i$ \\
$\hat{D}_i^m$ &~~ average difficulty between blocks $k_{i-1}^m$ and $k_i^m$ \\
$\mu$ &~~ maximum multiplicative increase in commitment over previous average \\ 
$\gamma$ &~~ maximum fraction of total hash rate allowed in Bootstraping state \\
$\tau_s$, $\tau_l$ &~~ short- and long-range KS test tolerances \\
$\Delta_n$ &~~ KS test statistic for window size $n$ \\
$p(\Delta_n)$ &~~ p-value for $\Delta_n$ \\
$\kappa$ &~~ miner cost tolerance, expressed as fraction of bond $b$ \\
\bottomrule
\end{tabular}
}

\end{document}